\documentclass[sigconf]{acmart}

\usepackage{adjustbox}
\usepackage{array,booktabs}
\usepackage{amsmath, amsfonts, amsthm}
\usepackage{amssymb}
\usepackage{bm}
\usepackage{graphicx}
\graphicspath{{./}{Charts/}}
\usepackage{subcaption}
\usepackage{tikz}
\usepackage{xspace}
\usepackage{algorithm}
\usepackage{todonotes}
\usepackage[noend]{algpseudocode}
\usepackage{hyperref}

\usepackage{booktabs} 

\copyrightyear{2019}
\acmYear{2019} 
\setcopyright{iw3c2w3}
\acmConference[WWW '19]{Proceedings of the 2019 World Wide Web Conference}{May 13--17, 2019}{San Francisco, CA, USA}
\acmBooktitle{Proceedings of the 2019 World Wide Web Conference (WWW '19), May 13--17, 2019, San Francisco, CA, USA}
\acmPrice{}
\acmDOI{10.1145/3308558.3313680}
\acmISBN{978-1-4503-6674-8/19/05}

\usepackage{balance}

\usetikzlibrary{decorations.pathreplacing}

\AtBeginDocument{}
\AtBeginDocument{\newtheorem{prop}[theorem]{Proposition}}
\AtBeginDocument{}
\AtBeginDocument{\newtheorem{defn}{Definition}}
\AtBeginDocument{}

\DeclareMathOperator*{\argmin}{arg\,min}
\DeclareMathOperator*{\argmax}{arg\,max}

\newcommand{\rand}{Random}
\newcommand{\ahead}{Greedy}
\newcommand{\gonz}{Gonzalez}
\newcommand{\greed}{Myopic}
\newcommand{\nvgreed}{Na\"ive Myopic}

\usepackage[draft,inline,nomargin,index]{fixme}
\fxsetup{theme=color,mode=multiuser,inlineface=\itshape,envface=\itshape}
\FXRegisterAuthor{sv}{asv}{\colorbox{gray!10!white}{\color{black}Suresh}}
\FXRegisterAuthor{sf}{asf}{\colorbox{magenta!10!white}{\color{black}Sorelle}} 
\FXRegisterAuthor{cs}{acs}{Carlos} 
\FXRegisterAuthor{bf}{abf}{\colorbox{blue!10!white}{\color{black}Ben}}
\FXRegisterAuthor{db}{adb}{\colorbox{green!10!white}{\color{black}danah}}
\FXRegisterAuthor{ab}{aab}{\colorbox{red!10!white}{\color{black}Ashkan}}

\begin{document}

\title{Gaps in Information Access in Social Networks}
\titlenote{This research was funded in part by the NSF under grants IIS-1633387,
  IIS-1633724, IIS-1513651, and IIS-1526379.}
\author{Benjamin Fish}
\affiliation{%
  \institution{Microsoft Research}
}
\email{benjamin.fish@microsoft.com}

\author{Ashkan Bashardoust}
\affiliation{
\institution{University of Utah}
}
\email{ashkanb@cs.utah.edu}

\author{danah boyd}
\affiliation{\institution{Data \& Society}}
\email{danah@datasociety.net}

\author{Sorelle A. Friedler}
\affiliation{%
  \institution{Haverford College}
}
\email{sorelle@cs.haverford.edu}

\author{Carlos Scheidegger}
\affiliation{%
  \institution{University of Arizona}
}
\email{cscheid@cscheid.net}

\author{Suresh Venkatasubramanian}
\affiliation{%
  \institution{University of Utah}
}
\email{suresh@cs.utah.edu}

\renewcommand{\shortauthors}{B. Fish et al.}

\begin{abstract}


  The study of influence maximization in social networks has largely ignored
  disparate effects these algorithms might have on the individuals contained in the
  social network.  Individuals may place a high value on receiving
  information, e.g.~job openings or advertisements for loans.  While
  well-connected individuals at the center of the network are likely to receive
  the information that is being distributed through the network, poorly
  connected individuals are systematically less likely to receive the
  information, producing a gap in access to the information between individuals.
  In this work, we study how best to spread information in a social network
  while minimizing this access gap.

  We propose to use the maximin social welfare function as an objective
  function, where we maximize the minimum probability of receiving the
  information under an intervention.  We prove that in this setting this welfare
  function constrains the access gap whereas maximizing the expected number of
  nodes reached does not.  We also investigate the difficulties of using the
  maximin, and present hardness results and analysis for standard greedy
  strategies.  Finally, we investigate practical ways of optimizing for the
  maximin, and give empirical evidence that a simple greedy-based strategy works
  well in practice.

\end{abstract}

%
%



\maketitle

\section{Introduction}
\label{sec:introduction}

Information flow in networks has been a subject of extensive study. Among the
many motivations for the study of how information propagates in a network has
been advertising (how can we spread information most effectively on a budget) and clustering (how do groups form and organize in a network).

One of the most important questions in this area is how to maximize influence in a social network.  Here the goal is to choose where to place initial sources of information so as to maximize the flow of information via word-of-mouth.  First formalized by Kempe, Kleinberg, and Tardos~\cite{kempe2003maximizing}, there has been a long series of work in the literature on influence maximization. 

However, this work has not typically focused on the impact that the information has on the individuals in the network.  
For example, one important application of information flow in networks is for recruitment.  Social networks like LinkedIn are increasingly used to provide
access to jobs and information that can greatly impact an individual's career
development.  Often just as important as the individuals themselves
are the connections \emph{between} individuals -- their social networks -- in making hiring decisions.  This is because information transmitted amongst social networks may accrue amongst the best-connected individuals in the network.  As the adage goes, ``it's not what you know, but who
you know.''  With more and more of our social life mediated through online
networks, the role that networks play in opening up opportunities is
increasingly important.  This includes not only recruitment, but also advertising and other kinds of marketing. 

However, network structure can create haves and have-nots in the game of
access. Insiders who are well-connected in the network have easier access to
relevant information about opportunities for advancement that can in turn lead
to even better connections. Outsiders who lack access to such information will
find it much harder to improve their network status. This access gap may lead to a form of inequality that is different from the traditional
forms of inequality based on class, race, gender, or other attributes, but nonetheless provides a significant challenge.

Thus, we are concerned with each individual's access to information and not just the number of people reached or the amount of information being distributed.  
How might we ensure that the
access gap in information is reduced? Rather than asking how far we can spread
information on a budget, we instead ask which people are getting the
information we're spreading.

\subsection{Our Work}

How can we formulate a notion of equitable access to information in a network, and how might we intervene in a network (on a budget) to minimize the gap in access to information?  In particular, we examine how best to add seeds (individuals who start with the information) to a network to minimize this gap in access.

We propose a new measure of access in a network.  In contrast to previous work that maximizes the average probability that an individual receives the information (\emph{max reach}), we instead propose to maximize the minimum probability.  
We formalize access as a \emph{social welfare function} that assigns a real value to the set of utilities received by the individuals, in this case the probabilities of receiving the information.  This allows us to evaluate the notions of access themselves:  we consider a notion of access to be better if interventions that optimally maximize that notion do not widen the access gap.  We show that every notion of access (amongst a wide class of such functions) does to some degree permit the access gap to increase in the worst case.  On the other hand, if the access gap increases between two groups of individuals after an intervention, we show that our proposed notion of access at least prohibits situations where the access does not increase at all for the group which started off with less access to the intervention.  Perhaps surprisingly, we show in Section~\ref{sec:unfairness} that a very large class of natural notions of access (including maximum reach) does not have this very basic prohibition.  We desire this because without such a prohibition, in the worst case there's nothing stopping interventions from creating one permanently and significantly advantaged group with access to information and one group without any such access, which we regard as blatantly undesirable.

We show that maximizing the minimum probability is NP-hard, hard even to approximate well, and moreover that a number of standard greedy strategies have asymptotically worst-possible approximation ratios.  Nonetheless, we show via experiments that a very simple greedy strategy performs well in practice:  namely, choose the seeds to be the vertices currently estimated as having the smallest probabilities of receiving the information. 
We also demonstrate that by using this strategy, we decrease the correlation between vertices' probability of receiving the information and their location in the network, indicating that our measure of access is not merely a proxy for (static) network structure. 

\paragraph{Limitations}
We recognize that asking to maximize the minimum probability of access to information ignores the fact that not all individuals in a network might need a particular piece of information. For example, a hiring ad should be spread widely, but only to candidates who are eligible, are in the right geographic areas, and have desirable qualifications. More generally, interventions to improve access to information might themselves cause feedback loops (both virtuous and vicious): our work does not consider those dynamics.  Nor does our work consider other notions of utility, like those that take into account the benefits of receiving the information more than once.  We leave study of these issues for future work.

~\\
In summary, our main contributions are as follows.
\begin{itemize}
\item We propose a new measure of information access in a network. We
  demonstrate that this measure captures certain axiomatically desirable
  properties of any notion of equal access, and further that existing notions
  including the well-studied \emph{maximum reach} concept do not. 
\item We investigate the problem of maximizing access theoretically,
  presenting hardness results as well as analysis of standard greedy strategies.
\item We do a comprehensive empirical evaluation of heuristics for
  achieving a high level of access, demonstrating that a greedy-based strategy is
  quite effective at improving equality of access in a network for a given
  budget of interventions. 
\end{itemize}

\subsection{Related Work}
\label{sec:disc-limit}

Granovetter's seminal work on the strength of weak ties~\cite{granovetter1977strength} first broached the idea that network position can confer advantages or disadvantages (including in hirings scenarios).  Indeed, weak ties can influence success in hiring and careers~\cite{granovetter1983strength}.  In an algorithmic setting, boyd, Levy, and Marwick~\cite{boyd2014networked} illustrate how modern social networks like LinkedIn might be vehicles for a more direct propagation of advantage and disadvantage.  In that light, our work, which focuses on how to mitigate such effects in the context of information access, falls into the paradigm explored by fairness-aware decision-making in which the goal is to design decision-making systems that ensure the end result is non-discriminatory to individuals or groups of individuals. Our work can be viewed as an attempt to quantify one aspect of \emph{social capital}, a notion introduced by Coleman \cite{coleman1988social} to capture how social standing within a system could be interpreted as a resource that has utility for an agent.  Recently, Benthall and Haynes~\cite{benthall2019racial} consider how to use a social network to define racial aspects of social standing, but don't consider interventions in the social network.

Rather than directly model an explicit fair goal for a decision in this setting, via assuming we have access to a sensitive feature like race on which we would focus our attention, we instead model the utility that each individual receives.  This formalizes how best to optimize for access to information without necessarily requiring \emph{equal} access.  While most of the literature in algorithmic fairness uses equality-based definitions \cite{Dwork12Fairness,Romei13Multidisciplinary,fish2016confidence,hardt2016equality,2015_kdd_disparate_impact,zafar2017fairness,Narayanan2018} (typically either group fairness or individual fairness), the welfare approach to fairness that we use is starting to become more popular.  For example, Heidari et al.~\cite{heidari2018fairness} propose a specific welfare function to use for classification and regression problems.  

Our choice of welfare function is based on axiomatic considerations: by determining which functions satisfy specific mathematical criteria used to model gaps in access. The resulting function that seeks to maximize the minimum probability of receiving information bears some resemblance to the \emph{difference principle} outlined by Rawls~\cite{rawls2009theory}, in that it seeks to intervene so as to provide benefit to the ``least-advantaged'', here interpreted as those with the least probability of access.  

Our work relies on a framework for information propagation that comes from the broad area of \emph{influence maximization}. Influence maximization seeks ways to spread information in a network efficiently using a small collection of \emph{seeds}. The typical measure of information spread used is the expected number of nodes that receive the information (the max reach measure).  While influence maximization assigns the same utility to an individual as we do, the welfare function in that setting is just the sum of the individual utilities.  This \emph{utilitarian} approach was initiated by Domingos and Richardson~\cite{richardson2002mining} and is formalized as a discrete optimization problem in Kempe, Kleinberg, and Tardos~\cite{kempe2003maximizing}.  There is also work into making this process faster~\cite{chen2009efficient, tang2014influence} or suitable for more general situations, where factors like pricing must be taken into account~\cite{arthur2009pricing}.  

A related body of algorithmic work~\cite{garimella2017reducing,matakos2017measuring,musco2018minimizing} posits that one way to decrease polarization in social networks is to connect people with opposing views by exposing them to new information.  Such work differs in focus and approach to modeling from this work because that work is concerned with poor connectivity between communities and we are concerned with individuals who are simply poorly connected.





\section{Definitions}\label{sec:definitions}

Let $G$ be a graph with $n$ nodes. To describe information flow in $G$ we will use a standard probabilistic model for how information travels -- the \emph{independent cascade} (IC) model~\cite{kempe2003maximizing}.
In this model, a node either possesses information or not.  A set of \emph{seed} nodes start out with the information, and information flow proceeds in rounds. Each newly informed node $v$ informs its neighbors $u$ in the next round i.i.d.~with probability of transmission $\alpha_{u,v}$.  Once a node is informed, it stays informed, and no longer passes on the message. In this work, we will use the IC model with a fixed probability $\alpha$ of transmission.

\paragraph{Welfare Functions}
\label{sec:welfare-functions}

In the IC model with parameter $\alpha$, we can associate with each vertex $v$ the probability $p_v$ that $v$ is informed after all information has been passed. We now define a social welfare function $\mu:[0,1]^n\rightarrow\mathbb{R}$ to represent how effectively information is spread:  it takes as input the probability of infection for each vertex, and outputs the overall welfare. 
\begin{defn}
The welfare of a set of vertices $V=\{v_1,\ldots,v_{|V|}\}$ in $G$ with seed set $S$ is 
$\mu_G(S,V) = \mu(p_{v_1},\ldots,p_{v_{|V|}})$.
If $V$ is all $n$ vertices, we abbreviate this as $\mu_G(S)$.  
 \end{defn}
\noindent When the graph is clear from context, we will omit the subscript $G$ and write $\mu(S,V)$ and $\mu(S)$ respectively.

Seed sets represent an \emph{intervention} in the information network. Thus, a primary goal in the study of information flow is to find a \emph{budgeted intervention}: a set of seeds $S_+$ of size no more than $k$ for a given graph $G$ (possibly with initial seeds $S$) with maximum welfare
\[S^* = \argmax_{\substack{S_+\cup S:\\|S|\le k}}\mu_G(S_+\cup S).\]
In other words, $S^*$ is the initial seeds $S$ along with a set of $k$ vertices which maximizes access for $G$.
Later, we will also consider the set of seeds that maximize welfare for a particular set of vertices:
\[S_V = \argmax_{\substack{S_+\cup S:\\|S|\le k}}\mu_G(S_+\cup S,V).\]

Kempe, Kleinberg, and Tardos~\cite{kempe2003maximizing} and subsequent work use as their welfare function \emph{reach}, the expected number of nodes reached.  In our notation, and normalizing to make it conveniently $[0,1]$-valued, this becomes the following:

\begin{defn}[Reach] $\mu_{\text{reach}}(S,V) = \frac{1}{|V|}\sum_{v\in V}p_v$.
\end{defn}

We can easily generalize this to a wider class of notions of welfare.  We consider generalized means:

\begin{defn}[$\phi$-mean] $\mu_{\phi}(S,V) = \left(\frac{1}{|V|}\sum_{v\in V} p_v^\phi\right)^{1/\phi}$.
\end{defn}
Note in the limit, this becomes the geometric mean for $\phi=0$, the minimum for $\phi=-\infty$, and the maximum for $\phi=+\infty$.  In other words,
$\mu_{-\infty}(S,V) = \min_{v\in V} p_v$.

We say that a function $\mu_G(S,V) = \mu(x_1,\ldots,x_m)$, each $x_i\in[0,1]$ representing the probability that a node $i$ receives the information, is \emph{monotonically increasing} if $\mu(x_1,\ldots,x_m) \ge \mu(x'_1,\ldots,x'_m)$ when $x_i \ge x'_i$ for all $i$.  A function $\mu$ is \emph{strictly monotonically increasing} if $\mu(x_1,\ldots,x_m) > \mu(x'_1,\ldots,x'_m)$ when $x_i \ge x'_i$ for all $i$ and in addition there is some $j$ such that $x_j > x'_j$.  $\mu$ is \emph{symmetric} if $\mu(x_1,\ldots,x_m)=\mu(x_{\sigma(1)},\ldots,x_{\sigma(m)})$ for all permutations $\sigma$.

In this work, we restrict our attention to symmetric, monotonically increasing welfare functions so that no vertex is privileged above the others and, all else equal, increasing an individual's probability of receiving the information is never undesirable.  The $\phi$-means are such functions.  Moreover, if a continuous welfare function satisfies four natural conditions (symmetry, strictly monotonically increasing, independence of unconcerned agents, and independence of common scale\footnote{Independence of common scale means that the ordering over alternatives should not change when multiplying each probability by a common positive factor, and independence of unconcerned agents means that the ordering should be independent of a probability that doesn't change, i.e.~if $\mu(x,x_1,\ldots,x_m) \ge \mu(x,x'_1,\ldots,x'_m)$, then $\mu(y,x_1,\ldots,x_m) \ge \mu(y,x'_1,\ldots,x'_m)$ for all $y$.
}) as a consequence of the Debreu-Gorman theorem~\cite{debreu1959topological, gorman1968structure} the only such welfare functions up to ordering over preferences are the $\phi$-means~\cite{heidari2018fairness, roberts1980interpersonal}, as long as all probabilities are non-zero.  In other words, at least in the case of connected undirected graphs, $\phi$-means are an extremely wide class of symmetric, monotonically increasing welfare functions, making them a natural class to examine.

\section{Gaps in Access}\label{sec:unfairness}

Optimizing a welfare function is a way to improve access to information in the aggregate. But our concern in this work is whether individuals or subgroups are being left behind in the process. Is it possible that even though an aggregate measure of information access is increasing, the gap in information access between groups is getting larger?
In this section, we will focus on evaluating welfare functions with respect to information access properties we would like to ensure.


We now define the \emph{access gap}, which captures how much better some individuals are doing than others.
\begin{defn}
The \emph{access gap} of a (non-trivial) partition $V,V'$ of the vertices of a graph $G$ with seed set $S$ under a welfare function $\mu$ is \[\mu(S,V)-\mu(S,V').\]  
\end{defn}

Note we only define the access gap over bipartitions, rather than arbitrary subsets.  This is to prevent the following situation:  Given a partition $V_1,V_2, V_3$ of $G$ and initial seed set $S$, suppose $\mu(S,V_1)=\mu(S,V_2)$ are both very large, but $\mu(S,V_3)$ is much smaller.  Consider $S^*$, the optimal seed set for this graph, and suppose now $\mu(S^*,V_1) > \mu(S^*,V_2) = \mu(S^*,V_3)$.  We now have a gap between the access of $V_1$ and $V_2$, but this gap was a by-product of significantly increasing the access of $V_3$.  Since this may well be desirable behavior, we preclude this situation by only considering gaps between bipartitions.


In particular, we want to know when the access gap increases.  We call this the \emph{rich getting richer} phenomenon.
\begin{defn}[Rich get richer]

In a graph $G$ with initial seeds $S$ under a welfare function $\mu$, we say that the \emph{rich get richer} if there is a (non-trivial) partition $V,V'$ where the optimal intervention $S^*$ satisfies
\[\mu(S^*,V')-\mu(S^*,V) > \mu(S,V') - \mu(S,V) > 0.\]
\end{defn}

Unfortunately, stopping the rich from getting richer in arbitrary graphs may be too much to hope for.  Even simple examples show that under many notions of welfare, including all $\phi$-means, the rich get richer.

\begin{figure}


  \includegraphics[width=0.5\columnwidth]{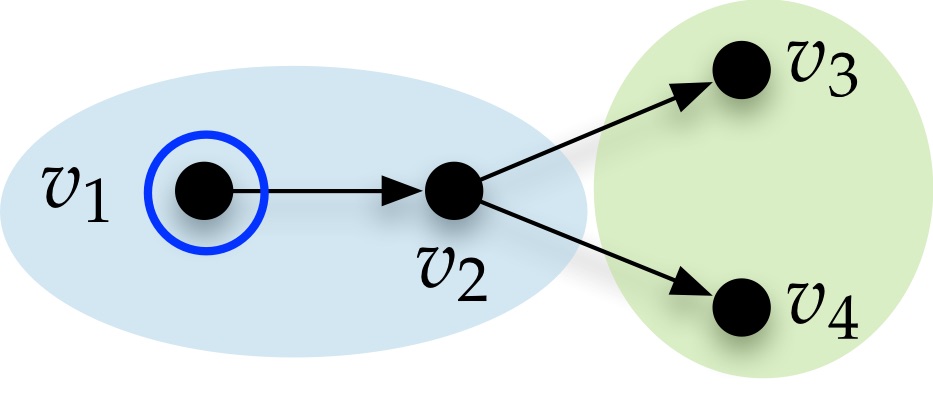}
\caption{Example showing that the rich can get richer under the optimal intervention.  If only one additional seed may be added, it is $v_2$ for any monotonic welfare measure.  Under this intervention, the access gap between $\{v_1,v_2\}$ and $\{v_3,v_4\}$ (the two colored sets) widens.}
\label{fig:rgr_counterexamp}
\end{figure}

\begin{prop}
\label{prop:richer}
Suppose $\mu$ is symmetric, increasing, and satisfies the following condition:
For any $x_1$, \ldots, $x_m$ in $[0,1]$, there is some $1\le \phi<\infty$ such that 
\[\min_{i} x_i  \le \mu(x_1,\ldots,x_m) \le \left(\frac{1}{m}\sum_{i=1}^m x_i^\phi\right)^{1/\phi}.\]
Then under $\mu$, when $0 < \alpha < \frac{1}{2\phi}$, there exists a graph and initial seed set where the rich get richer.
\end{prop}

Note that the upper bound in this third condition is easy to satisfy; it suffices that $\mu(x_1,\ldots,x_m)$ is strictly less than $\max_i x_i$ when not all of the $x_i$ are equal to each other.  In addition the assumption that $\phi\ge 1$ is only assumed for the sake of convenience, since $\left(\frac{1}{m}\sum x_i^\phi\right)^{1/\phi}$ is monotonic in $\phi$.


\begin{proof}
Consider the example graph $G$ in Figure~\ref{fig:rgr_counterexamp} and suppose $0<\alpha<1$.  Let $V=\{v_3,v_4\}$ and $V'=\{v_1,v_2\}$.  Note that $p_{v_1} = 1$, $p_{v_2} = \alpha$, and $p_{v_3} = p_{v_4} = \alpha^2$.  Then $\mu(S,V) = \mu(p_{v_3},p_{v_4}) = \mu(\alpha^2, \alpha^2) = \alpha^2$.  Yet $\mu(S,V') = \mu(1,\alpha) \ge \alpha > \alpha^2$, so we have $\mu(S,V') > \mu(S,V)$.

What is the optimal seed to add?  If we add $v_2$ to the seeds, then we have $p_{v_1} = p_{v_2} = 1$ and $p_{v_3} = p_{v_4} = \alpha$.  Otherwise, if we add $v_3$ to the seeds, then $p_{v_1} = p_{v_3} = 1$, $p_{v_2} = \alpha$, and $p_{v_4} = \alpha^2$.  Note $\mu(1,1,\alpha,\alpha) \ge \mu(1,\alpha,1,\alpha^2)$ by symmetry and monotonicity, so without loss of generality the optimal modification is to make $v_2$ a seed.  Then it is easy to calculate $\mu(S^*,V')-\mu(S^*,V) = \mu(1,1) - \mu(\alpha^2,\alpha^2) = 1-\alpha^2$.  Thus we have the rich getting richer if $1-\alpha > \mu(1,\alpha)-\alpha^2$.  But $\mu(1,\alpha) \le \left(\frac{1+\alpha^\phi}{2}\right)^{1/\phi}$, so it suffices to show that
$\frac{1+\alpha^\phi}{2} < \left(1-\alpha+\alpha^2\right)^\phi$.  Then since $\phi\ge 1$ and $0 < \alpha < \frac{1}{2\phi}$,

\[\frac{1+\alpha^\phi}{2} \le \frac{1+\alpha}{2} < 1-\phi\alpha+\phi\alpha^2 \le (1-\alpha+\alpha^2)^\phi.\]
\end{proof}

Proposition \ref{prop:richer} holds for all $\phi$-means for $\phi<\infty$.  We will show in Section~\ref{sec:imbalance} that the rich get richer not only for the $+\infty$-mean but a whole other class of welfare functions as well (a consequence of Proposition~\ref{prop:strict_increase_imbalanced}).
Given this, keeping the rich from getting richer appears to be too much to hope for.

\subsection{$k$-imbalance}\label{sec:imbalance}
If we can't keep the rich from getting richer in the worst case, what can we prevent?  A particularly concerning case of the rich getting richer is when the access of the worse-off group $V$ doesn't improve at all.  That is, a case where $\mu(S,V') >\mu(S,V)$ under the initial seeds $S$ and the rich get richer, but for the set of seeds $S^*$ that maximize welfare $\mu(S^*,V) \le \mu(S,V)$.  This might not be so bad if the only way to improve the access of $V$ is to increase the access of $V$ to the point where it is even higher than that of $V'$, so that $V'$ becomes the worse-off group.  On the other hand, this situation becomes particularly egregious when in addition $\mu(S_V,V) \le \mu(S,V')$, i.e.~the optimal improvement for $V$ still does not improve the access of $V$ to the point where it is larger than the access that $V'$ started out with prior to intervention (recall that $S_V$ -- defined in Section~\ref{sec:definitions} -- is the seed set that maximizes reach for $V$).  If this can happen when adding $k$ seeds, we will call $\mu$ $k$\emph{-imbalanced}.  That is, $k$-imbalance is a particularly egregious form of the rich getting richer.  If $\mu$ is not $k$-imbalanced for any $k>0$, we will call it \emph{balanced}.  

We believe that balance is a natural desideratum because it prevents interventions from never helping the worse-off group at all.  
Stronger versions of preventing disparity in access may still be preferred, like avoiding the rich from getting richer, so balance may only represent a necessary but not sufficient condition for preventing disparity.  
In this section, we show a wide class of $\mu$ are $\Omega(n)$-imbalanced, but that $\mu_{-\infty}$ is balanced.



\begin{defn}[$k$-imbalance]
A welfare function $\mu$ is $k$-\emph{imbalanced} if there exists a graph $G$ with initial seed set $S$ and partition of the vertices $V$ and $V'$ where the optimal intervention $S^*$ and optimal intervention for $S_V$ under the addition of no more than $k$ seeds satisfies the following:
\begin{enumerate}
\item $\mu(S,V) < \mu(S_V,V)$ \emph{(There is a set of seeds to add that improves the access of $V$.)}
\item $\mu(S_V,V) \le \mu(S,V')$ \emph{(Not only does $V'$ start off with more access than $V$, but $V'$ starts off with more access than $V$ can possibly achieve.)}
\item $\mu(S^*, V') > \mu(S,V')$  \emph{(The access of $V'$ improves.)}
\item $\mu(S^*, V) \le \mu(S,V)$ \emph{(The access of $V$ does not improve.)}
\end{enumerate}
\end{defn}

In other words, a welfare function is imbalanced if
\[\mu(S^*,V) \le \mu(S,V) < \mu(S_V,V) \le \mu(S,V') < \mu(S^*,V').\]

Note that it is immediate that if $\mu$ is $k$-imbalanced for any $k>0$, then the rich get richer under $\mu$.  As $k$ increases, it should be the case that it becomes more difficult to find examples of imbalance, as it is harder to avoid improving the access of $V'$.  Nonetheless, we can show that a wide class of welfare functions, including reach, is $\Omega(n)$-imbalanced:
\begin{prop}\label{prop:strict_increase_imbalanced}
Suppose $\mu$ is symmetric and strictly increasing.  Then $\mu$ is $\Omega(n)$-imbalanced.
\end{prop}
\begin{proof}
It suffices to consider the simplest case:  when there is no communication, i.e.~$G$ is the disjoint graph of $n$ vertices.  $V$ and $V'$ will each be exactly half of the vertices (for $n$ even).  The initial seed set $S$ will be entirely contained in $V'$ and will be size $n/4$.  Now we will add an additional $n/4$ seeds.  Note first that since $\mu$ is symmetric, each of the vertices (with the exception of the initial seeds) are identical.  So $S_V$ is any set of $n/4$ additional seeds in $V$:  each additional seed must improve the welfare of $V$ because $\mu$ is strictly increasing.  But in this case, $V$ and $V'$ become identical, so we have $\mu(S,V) < \mu(S_V,V) \le \mu(S,V')$.  But by symmetry, the optimal seeds to add can be any $n/4$ vertices, in which case we can assume they are all in $V'$.  Thus the welfare of $V$ does not increase while the welfare of $V'$ does.
\end{proof}

It turns out that balance is a useful definition, insomuch as it is actually possible to achieve.
\begin{prop}
$\mu_{-\infty}$ is balanced.
\end{prop}
\begin{proof}
Suppose $\mu_{-\infty}$ is imbalanced, witnessed by some partition $V,V'$ of $G$ and initial seed set $S$.  Recall imbalance implies that $\mu_{-\infty}(S,V) < \mu_{-\infty}(S,V')$.  Then by definition of $\mu_{-\infty}$, the vertex $v$ with minimum probability is in $V$, i.e.~$\mu_{-\infty}(S) = \mu_{-\infty}(S,V)$.  Remember $S^*$ maximizes the minimum probability, and $\mu_{-\infty}(S_V,V) > \mu_{-\infty}(S,V)$, so there is at least one graph that increases that minimum probability, which in turn means that $S^*$ does as well.  Thus $\mu_{-\infty}(S^*,V) > \mu_{-\infty}(S,V)$, a contradiction.
\end{proof}

On the other hand, $\mu_{-\infty}$ is a special case, and every other $\phi$-mean is maximally imbalanced:  there exists a graph, initial seed set, and partition of the vertices that verifies the other $\phi$-means are imbalanced. 


\begin{prop}\label{prop:pmean_imbalanced}
For $\phi > -\infty,\alpha<1, \mu_\phi$ is $\Omega(n)$-imbalanced. 
\end{prop}
\begin{proof}
If $\phi=+\infty$, so $\mu_\phi$ is the maximum probability, then as soon as the graph has at least one seed, then $\mu_\phi(S)=1$, and any added seeds after that don't change the value, so $\mu_\phi$ is trivially $\Omega(n)$-imbalanced.  Otherwise, if $\phi>0$, $\mu_\phi$ is strictly increasing, and from Proposition~\ref{prop:strict_increase_imbalanced} we know it is $\Omega(n)$-imbalanced.  And if $\phi\le 0$, then $\mu_\phi$ is strictly increasing once all probabilities are non-zero, at which point we use a similar tactic to when $\mu$ is strictly increasing, except we will need a connected graph.  We will use the star graph, with one central vertex the seed, and all other vertices connected to that seed.  In addition there will be some $n/2-1$ additional seeds, all in $V'$, which consists of those seeds, the central seed, plus $n/2$ more vertices.  $V$ will be the other $n$ vertices, so that $G$ is $2n$ nodes.  Our goal will be to add an additional $n/2$ seeds.  Remember, since $G$ is connected (all vertices have non-zero probability) $\mu_\phi$ is strictly increasing.  Then the optimal graph for $V$ is to add all $n/2$ additional seeds to $V$, in which case we have $n/2$ vertices with probability 1 and $n/2$ vertices with probability $\alpha$.  But $V'$ in $G$ is exactly the same, so we have $\mu(S,V) < \mu(S_V,V) \le \mu(S,V')$.  However, all non-seeds are isomorphic, so we may assume all $n/2$ new seeds are added to $V'$.
\end{proof}

We note that one could consider many variations of $\phi$-means, including replacing mean with median, maximum with minimum, etc. These variations do not affect the results that we present here.  We defer a detailed analysis of these variations to the full version of the paper. 



\section{Maximin access}

The previous section established $\mu_{-\infty}$ as a better access measure than others, at least when it comes to achieving balance. We now study the problem of 
maximizing $\mu_{-\infty}$, which we call the \emph{maximin access} problem.  We start by showing that this is NP-hard even to approximate well.  


\begin{theorem}\label{thm:maxmin_hardness}
Suppose $\alpha < \frac{\sqrt{5}-1}{2}$.  Then choosing $k$ seeds to maximize min access is NP-hard.  In this case, the maximin access cannot be approximated better than $O(\alpha)$ and if furthermore $\alpha = O(1/n)$ then the maximum cannot be attained efficiently without an additional $O(\ln n)$ factor seeds.
\end{theorem}
\begin{proof}
We reduce from {\sc Set Cover}, where an instance is defined by a collection of subsets $S_1,\ldots, S_m$ over a ground set $U=\{x_1,\ldots,x_n\}$ and an integer $k$, and the decision problem is whether or not there is a collection of $k$ subsets whose union is $U$.  Further, we can assume $k < n < m$.  Given such an instance, we construct a directed graph (example showed in Figure~\ref{fig:np_hardness}).  We start with the natural directed bipartite graph corresponding to a set cover instance, where there is a vertex $i$ corresponding with each set $S_i$ and a vertex $j$ corresponding with each element $x_j$.  There is a directed edge from $i$ to $j$ whenever $x_j$ is contained in $S_i$.  We then add a single extra vertex $v$ and directed edges from $v$ to each vertex $i$ corresponding with one of the sets, and ask to maximize the minimum probability by adding $k+1$ seeds.  

\begin{figure}
\centering
 \tikzstyle{vertex}=[circle,fill=black,minimum size=10pt,inner sep=0pt]
 \tikzstyle{circled} = [draw, blue, thick, circle, minimum size=.55cm]
 \tikzstyle{edge}=[<-,color=black,line width=1.5]
 \tikzstyle{edgeback}=[->,color=black,line width=1.5]
 \tikzstyle{edgelabel}=[color=black]
%
%
%
%
       
 \begin{tikzpicture}[scale=0.8, auto]

 \node[circled] () at (0,0){};
 \node[vertex] (v) at (0,0) [label=left:$v$]{};


 \node[circled] () at (-1,-1){};
 \node[vertex] (s1) at (-1,-1) [label=left:$S_1$]{}
 	edge[edge] node {} (v);
 \node[vertex] (s2) at (0,-1) [label=right:$S_2$]{}
 	edge[edge] node {} (v);
\node[vertex] (s3) at (1,-1) [label=right:$S_3$]{}
 	edge[edge] node {} (v);
	
 \node[vertex] (x1) at (-1,-2) [label=left:$x_1$]{}
 	edge[edge] node {} (s1);

 \node[vertex] (x2) at (0,-2) [label=right:$x_2$]{}
   	edge[edge] node {} (s2)
 	edge[edge] node {} (s1);
 \node[vertex] (x3) at (1,-2) [label=right:$x_3$]{}
  	edge[edge] node {} (s2)
 	edge[edge] node {} (s3);

       \end{tikzpicture}
\caption{Corresponds with the set cover problem `Is there $k=1$ set among $S_1=\{x_1,x_2\}$, $S_2=\{x_2,x_3\}$, $S_3=\{x_3\}$ that cover all elements?'  Note that since the answer is no, then the minimum probability is no more than $p_{x_3} = 1-(1-\alpha^2)^2$.}
\label{fig:np_hardness}
\end{figure}
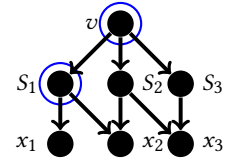
 
Since $v$ has in-degree zero, in order for the maximin access to be greater than zero, $v$ must be chosen as a seed.  In this case, since $k<m$, regardless of which seeds are chosen, there is some set $S_i$ such that $p_i = \alpha$.  Therefore the maximum min access is no more than $\alpha$.  Without loss of generality, no vertex corresponding to an element $x_j$ need be chosen as a seed.  Otherwise, the seed may be moved to any vertex corresponding with a set $S_i$ such that $x_j\in S_i$.  The maximin access cannot go down, because we still have $p_j\ge \alpha$.

If there is a set cover, then the maximum min access is at least $\alpha$:  choose the vertices corresponding to the cover for the seeds (plus $v$), in which case $p_v=1$, $p_i\ge \alpha$ because they are either seeds or distance one from $v$, and $p_j \ge \alpha$, because they are distance one from a seed.  If there is no set cover, then there is no way to choose the seeds amongst the $S_i$ such that all vertices are within distance one from a seed.  Assume that every element $x_j$ is contained in at most two subsets amongst the $S_i$ (this is now the {\sc Vertex Cover} problem, an NP-hard special case of {\sc Set Cover}).  So there must be some $p_j$ such that $p_j \le 1-(1-\alpha^2)^2$.  Thus when $\alpha > 1-(1-\alpha^2)^2$, i.e.~$\alpha < \frac{\sqrt{5}-1}{2}$, any algorithm that maximizes the min access chooses the set cover if there is one.  So any algorithm that has an approximation ratio strictly better than $\frac{1-(1-\alpha^2)^2}{\alpha} = O(\alpha)$ must in fact be exact, and therefore also find the set cover.

Even in the general case of {\sc Set Cover}, we can still distinguish between when there is and is not a set cover:  The existence of a set cover still means the maximin probability is at least $\alpha$, while the lack of a set cover implies there is at least one vertex with probability no more than $1-(1-\alpha^2)^m$, which is upper-bounded by $\alpha$ when $\alpha = 1/m$.  Therefore, since set cover is $O(\ln n)$-inapproximable, we cannot approximate the best $k$ seeds to add without an additional $O(\ln n)$-factor seeds.
\end{proof}

Moreover, if we can find the seeds that maximize the minimum probability, even approximately, we can boost this result to also compute the minimum probability itself approximately.  This serves as additional evidence that this problem is hard, as there is no known method to even approximately compute the minimum probability.
\begin{prop}
If there is an $\alpha^{f(n)}$-approximation algorithm for maximin access, there is an $\alpha^{2f(n)+2}$-approximation for the minimum access of a vertex in a graph $G$ given a seed $s$.  That is, if the minimum access is $p_{\text{min}}$ in $G$, then we can given an estimate $\hat{p}$ such that
\[\alpha^{2f(n)+2}p_{\text{min}} \le \hat{p} \le (1/\alpha)^{2f(n)+2} p_{\text{min}}.\]
\end{prop}
\begin{proof}
Given an instance $(G,s,\alpha)$, we construct a graph $G'$ similar to the one in Figure~\ref{fig:dist_is_bad}.  (We may assume that $G$ is connected.)  If the diameter of $G$ is $\ell$, add to $G$ a simple undirected path of length $\ell$ starting from $s$, and call it $G'$.  Call the end of this path $v$.  In $G$, $p_{\text{min}}\ge \alpha^\ell$, which means that if we compute the single optimal seed in $G'$, it must be on the path from $s$ to $v$.

Define $x$ so that $\alpha^{\ell-x} = \alpha^x\cdot p_{\text{min}}$, i.e.
$x = \ell/2 - \frac{\log(1/p_{\text{min}})}{2\log(1/\alpha)}$.  Then the optimal placement for a seed is at distance $k$ from $s$, where $\lfloor x\rfloor \le k \le \lceil x \rceil$, because for any $k$ we have $p'_v = \alpha^{\ell-k}$ and $p'_{\text{min}} = \alpha^k\cdot p_{\text{min}}$, where $p'$ denotes probabilities in $G'$.

Suppose that we have a $(1/\alpha)^{f(n)}$-approximation algorithm for maximin access, and it chooses some seed distance $k'$ from $s$ (we may assume that the seed is on the simple path, otherwise we may always choose $k'=0$).  Since it is a $(1/\alpha)^{f(n)}$-approximation on a simple path, $k'$ must be within $f(n)$ of $k$.  Now we can approximate $p_{\text{min}}$ using $k'$ as an estimate of $k$:  We estimate it as $\hat{p} = \alpha^{\ell-2k'}$.

Then $\alpha^{\ell-2k'} \le \alpha^{\ell-2(k+f(n)} \le \alpha^{\ell-2(x+1+f(n))}$, and likewise $\alpha^{\ell-2k'} \ge \alpha^{\ell-2(x-1-f(n))}$, so this is within $\alpha^{2f(n)+2}$ of $p_{\text{min}} = \alpha^{\ell-2x}$.

\end{proof}

\subsection{Maximin algorithms}

The above results imply that it is hard to maximize $\mu_{-\infty}$ even 
approximately.  Nonetheless, Theorem~\ref{thm:maxmin_hardness}  still leaves
open the possibility of an $\alpha^{c}$-approximation (for fixed number of seeds
and $c>1$).  In this section, we present the heuristics we will use, along with
a few baselines.  We will show in Section~\ref{sec:appx_ratios} that,
unfortunately, these natural heuristics have a worst-possible approximation
ratio (a ratio exponential in $n$).  These results do not preclude good
performance in practice, which we discuss in
Section~\ref{sec:experiments}. 



Making our task yet more challenging is that, unlike maximizing reach~\cite{kempe2003maximizing}, maximin is not a submodular objective.\footnote{This can be seen using the construction in the proof of Proposition~\ref{prop:ahead_is_bad}, starting with one seed in the center of a simple path.  Adding one additional seed then does nothing, but adding two seeds increases the minimum probability.}  Nonetheless, it is natural to try a greedy approach, where in each iteration, we add to the seeds the vertex that maximizes the objective function.  We refer to this heuristic as \textbf{\ahead} (Algorithm~\ref{alg:greedy}).  To do this, we use the simple approach of estimating each probability $p_v$ for every possible vertex to add to the seed set.  (See below for details on how we estimate these probabilities.) 
\begin{algorithm}
\caption{\ahead}\label{alg:greedy}
\begin{algorithmic}[1]
\Require{Graph $G$, initial seed set $S$, number of seeds to add $k$}
\For{$k$ iterations}
	\ForAll{$j\not\in S$} 
		\State $prob \gets \text{ProbEst}(G,S\cup\{j\})$ \Comment{Algorithm~\ref{alg:mc}}
		\State $nextMin[j] \gets \min_{i}{prob[i]}$ \Comment{The minimum probability when the seeds are $S\cup\{j\}$}
	\EndFor
	\State $v \gets \argmin_{j}{nextMin[j]}$ 
	\State add $v$ to $S$
\EndFor
\State \textbf{return} $S$

\end{algorithmic}
\end{algorithm}
We contrast this approach to the faster approach, which we will call \textbf{\greed} (Algorithm~\ref{alg:myopic}), whereby we instead in each round choose the vertex with the currently smallest probability as the new seed, without actually evaluating the new value of the objective function.
\begin{algorithm}
\caption{\greed}\label{alg:myopic}
\begin{algorithmic}[1]
\Require{Graph $G$, initial seed set $S$, number of seeds to add $k$}
\State $k' \gets k$
\If{$|S| = 0$}
  \State Initialize $S$ as the vertex with the highest degree
  \State $k' \gets k-1$
\EndIf
\For{$k'$ iterations}
\State $prob \gets \text{ProbEst}(G,S)$ \Comment{Algorithm~\ref{alg:mc}}
\State $v \gets \argmin_{i}{prob[i]}$ \Comment{pick node with min probability}
\State add $v$ to $S$
\EndFor
\State \textbf{return} $S$
\end{algorithmic}
\end{algorithm}

We also consider a na\"ive variation (\textbf{\nvgreed}, Algorithm~\ref{alg:nm}) which, instead of proceeding in rounds, given initial estimates for the probabilities, picks for the seeds the $k$ vertices with the smallest probabilities.

\begin{algorithm}
\caption{\nvgreed}\label{alg:nm}
\begin{algorithmic}[1]
\Require{Graph $G$, initial seed set $S$, number of seeds to add $k$}
\State $k' \gets k$
\If{$|S| = 0$}
  \State Initialize $S$ as the vertex with the highest degree
  \State $k' \gets k-1$
\EndIf
\State $prob \gets \text{ProbEst}(G,S)$ \Comment{Algorithm~\ref{alg:mc}}
\State Add to $S$ the $k'$ vertices $i\not\in S$ with smallest probability $prob[i]$
\State \textbf{return} $S$
\end{algorithmic}
\end{algorithm}

So far, we have omitted how to estimate the probabilities for each vertex.  Unfortunately, computing the probability $p_v$ for each vertex exactly is \#P-hard~\cite{provan1983complexity}.  Even computing probabilities of receiving the information with a guaranteed approximation ratio is a long-standing open problem~\cite{karger1999randomized}.  So in this paper, we use a Monte Carlo method, simulating the IC model a fixed number of times, and estimating the probabilities for each vertex as the percent of times the information reached that vertex under the simulations (Algorithm~\ref{alg:mc}).  Of course, this requires having at least one seed, which is not the case in the first round of \greed\ and \nvgreed.  So we always choose the first seed as vertex with the highest degree.  This approach for dealing with the first round, as well as estimating the probabilities, provides a simple way to compare these heuristics.  As such, for the experiments we also choose the first seed as the highest degree vertex for the \ahead\ heuristic as well, again to simplify comparison.  We leave for future work other approaches for these issues.

\begin{algorithm}

\caption{ProbEst (Monte Carlo probability estimation)}\label{alg:mc}
\begin{algorithmic}[1]
\Require{Graph $G$, seed set $S$}
\Ensure{$\alpha$, Number of simulation rounds $R$}
\State Initialize $hits[i] \gets 0$ for each $i$ a vertex of $G$
\For{$R$ iterations} \Comment{Simulate the IC model $R$ times}
    \ForAll{$i\in S$}
    	\State $hits[i]{+}{+}$ \Comment{$hits[i]$ is the number of times $i$ has received the information}
    \EndFor
    \State $activeQueue \gets S$ \Comment{Keep track of which vertices are currently active}
    \While{$activeQueue$ non-empty}
    	\State Dequeue $i$ from $activeQueue$
	\ForAll{neighbors $j$ of $i$}
		\State $transmit \gets$ True with probability $\alpha$, else False
		\If{$j$ has not been in $activeQueue$ and $transmit$}
			\State $hits[j]{+}{+}$
			\State Enqueue $j$ to $activeQueue$
		\EndIf
	\EndFor
	
    \EndWhile
\EndFor
\State $prob[i] \gets hits[i] / R$
\State \textbf{return} $prob$

\end{algorithmic}
\end{algorithm}

%


An alternative approach that avoids estimating probabilities is to pick seeds
that are far from each other, under the intuition that a node far away from the
current seeds is likely to have a small $p_i$ and therefore should be picked as
the next seed. The resulting heuristic
is to pick in each round the node that is furthest from the current set of seeds
as the next seed; we call this heuristic \textbf{\gonz} because of its
resemblance to the well-known algorithm for $k$-center
clustering~\cite{gonzalez1985clustering}.


One could choose other proxies for the utility $p_v$ such as nodes of
low degree (or high degree), or nodes that do not contain seeds in a fixed
radius ball around them. In our experiments with these heuristics, they were dominated in both quality and
performance by the ones mentioned above, and  we will not discuss them further.

\subsection{Approximation ratios of maximin algorithms}\label{sec:appx_ratios}

We now show that \greed, \nvgreed, \ahead, and an exact version of \gonz~all
have approximation ratios that are exponential in $n$, even if we assume the
probabilities required by \greed, \nvgreed, and \ahead\ can be estimated exactly.  This is to emphasize that their poor behavior in the worst case doesn't just stem from the difficulty of approximating the probabilities given a seed set, but the heuristics themselves. 

\subsubsection{\greed~and \nvgreed}
Note that in the case $k=1$, \greed\ and \nvgreed~are identical algorithms.  Thus we can show that in this case, both algorithms behave poorly in the worst case, even in the non-trivial case when we start with at least one initial seed.

\begin{prop}\label{prop:greed_is_bad}
Given a graph and non-zero initial seed set, choosing as the seed with smallest $p_v$ yields a solution with approximation ratio no better than $O(\alpha^{n})$.
\end{prop}
\begin{proof}
Consider the graph $G$ depicted in Figure~\ref{fig:greed_is_bad}.  If we are allowed to add only a single additional seed besides the initial seed set $\{s\}$, then this algorithm will choose to add either $v_1$ or $v_2$, because in $G$ they minimize $\min_v p_v$, where $p_{v_1}=p_{v_2}=\alpha^{\ell+1}$.  But since we can only reach one of the two, we still have $\min_v p_v = \alpha^{\ell+1}$.  But the optimal vertex to add to the seed set is $t$, where now $\min_v p_v \ge \alpha^2$.  Then we get an approximation ratio no better than $O\left(\frac{\alpha^{\ell+1}}{\alpha^{2}}\right) = O\left(\alpha^n\right)$. 
\end{proof}


\begin{figure}
\centering
\tikzstyle{vertex}=[circle,fill=black,minimum size=10pt,inner sep=0pt]
\tikzstyle{circled} = [draw, blue, thick, circle, minimum size=.55cm]
\tikzstyle{edge}=[<-,color=black,line width=1.5]
\tikzstyle{edgeback}=[->,color=black,line width=1.5]
\tikzstyle{edgelabel}=[color=black]
\begin{tikzpicture}[scale=0.8, auto]
    \node[circled] at (0,0) [label=below:$s$]{};
    \node[vertex] (c0) at (0,0) {};
    \node[vertex] (c1) at (1,0) {}
        edge[edge] node {} (c0);
    \node[vertex] (c2) at (2,0) {}
        edge[edge] node {} (c1);
    \node[vertex] (c3) at (3,0) {};
    \node[vertex] (c4) at (4,0) [label=below:$t$]{}
    	edge[edge] node {} (c3)
	edge[edgeback,bend left] node {} (c0)
	edge[edgeback,bend left] node {} (c1)
	edge[edgeback,bend left] node {} (c2)
	edge[edgeback,bend left] node {} (c3);
    \node[vertex] (c5) at (5,1) [label=right:$v_1$]{}
    	edge[edge] node {} (c4);
    \node[vertex] (c6) at (5,-1) [label=right:$v_2$]{}
    	edge[edge] node {} (c4);

    \draw [decorate,decoration={brace,amplitude=8pt},xshift=2pt,yshift=8pt, line width = 1]
    (0,0) -- (4,0) node [black,midway,yshift=8pt] {$\ell$};

    \node[draw=none] at (2.525,0) {$\cdots$};
\end{tikzpicture}
\caption{$G$}
\label{fig:greed_is_bad}
\end{figure}


\subsubsection{\ahead}
We now consider what happens if \ahead~ is used to choose the $k$ seeds.  One problem with \greed~was that, as demonstrated via Figure~\ref{fig:greed_is_bad}, choosing the vertex with the smallest probability ignores the actual objective function (which in that example is maximized by choosing vertex $t$).  What happens when we attempt to maximize the actual objective function?  Again, we assume that for any seed set we are given the exact probabilities instead of approximate probabilities, which we refer to as a \emph{probability oracle}.

\begin{prop}\label{prop:ahead_is_bad}
\ahead, even with a probability oracle, has an approximation ratio no better than $O(\alpha^{n/6})$.
\end{prop}
\begin{proof}
Consider the simple undirected path of length $n$, with no initial seeds, where we may add $k=2$ seeds.  The greedy algorithm, in the first iteration, must choose the central vertex (assume $n$ is even).  In the second iteration, no vertex can increase the minimum probability, so the minimum probability is $\alpha^{n/2}$.  However, the optimal minimum probability is much larger:  If the two seeds trisect the path so that they are $n/3$ apart, then no vertex is distance more than $n/3$ from a seed, in which case the minimum probability is at least $\alpha^{n/3}$.
\end{proof}

\subsubsection{Minimax distance}

\gonz~is a heuristic to minimize the maximum distance of any vertex from a seed.  One motivation behind this algorithm is that in Figure~\ref{fig:greed_is_bad}, adding an edge from $s$ to $t$ in $G$ takes care of the issues found with \greed\ by ensuring that all vertices have distance no more than two from the seed.  In general, minimizing the maximum distance exactly is difficult, but even if we could do so, this approach still has a bad approximation ratio.


To show this, we construct a (sparse, max degree two) graph where nonetheless
the vertex $t$ furthest away from the seed still has a relatively high
probability of receiving the information.  This is the case for $H_\ell$, shown
in Figure~\ref{fig:dist_big_prob}, that's sufficiently sparse but $p_t$ is
large.

\begin{lemma}
The probability of $t$ being infected in $H_\ell$, where each edge has weight $\alpha = 1/2$, is $p_t=\Theta(1/\ell)$.
\end{lemma}
 \begin{proof}
 Denote by level $k$ the vertices distance $k$ from $s$, and by symmetry, the probability of being infected at that level $p_k$.  We want to calculate $p_\ell$.  Note $p_0=1$ and $p_{k+1} = 1 - (1-\alpha p_{k})^2 = p_k - \frac{1}{4}p_{k}^2,$ a variant of the logistic map.

 Then $\frac{1}{p_{k+1}} = \frac{1}{\frac{1}{4}p_k(4-p_k)} = \frac{1}{p_k} + \frac{1}{4-p_k}$.  Note $1/4 \le \frac{1}{4-p_k} \le 1/3$.  Unwinding the recurrence, we get $\frac{1}{p_{k+1}} = \frac{1}{p_0} + \sum_{j=0}^k \frac{1}{4-p_j}$, and in particular we have $\frac{1}{p_0} + \frac{k+1}{4} \le \frac{1}{p_{k+1}} \le \frac{1}{p_0} + \frac{k+1}{3},$
 i.e.~$\frac{1}{p_k} = \Theta(k)$.
 \end{proof}

\begin{figure}
\centering
\tikzstyle{vertex}=[circle,fill=black,minimum size=10pt,inner sep=0pt]
\tikzstyle{circled} = [draw, blue, thick, circle, minimum size=.55cm]
\tikzstyle{edge}=[<-,color=black,line width=1.5]
\tikzstyle{edgeback}=[->,color=black,line width=1.5]
\tikzstyle{edgelabel}=[color=black]
\begin{tikzpicture}[scale=0.8, auto]
    \node[circled] at (0,0) [label=below:$s$]{};
    \node[vertex] (c0) at (0,0) {};
    \node[vertex] (c11) at (1,0) {}
        edge[edge] node {} (c0);
    \node[vertex] (c12) at (1,-1) {}
        edge[edge] node {} (c0);
    \node[vertex] (c21) at (2,0) {}
    	edge[edge] node {} (c11)
        edge[edge] node {} (c12);
    \node[vertex] (c22) at (2,-1) {}
        edge[edge] node {} (c11)
        edge[edge] node {} (c12);
    \node[vertex] (c31) at (3,0) {};
    \node[vertex] (c32) at (3,-1) {};
    \node[vertex] (c41) at (4,0) {}
    	edge[edge] node {} (c31)
        edge[edge] node {} (c32);
    \node[vertex] (c42) at (4,-1) {}
        edge[edge] node {} (c31)
        edge[edge] node {} (c32);
    \node[vertex] (c5) at (5,0) [label=right:$t$]{}
        edge[edge] node {} (c41)
        edge[edge] node {} (c42);

    \draw [decorate,decoration={brace,amplitude=8pt},yshift=8pt, line width = 1]
    (0,0) -- (5,0) node [black,midway,yshift=8pt] {$\ell$};

    \node[draw=none] at (2.525,-0.5) {$\cdots$};
\end{tikzpicture}
\caption{$H_\ell$}
\label{fig:dist_big_prob}
\end{figure}

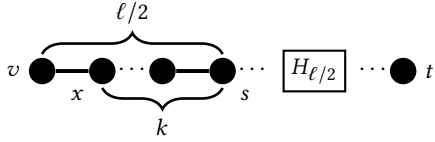
\begin{figure}
\centering
\tikzstyle{vertex}=[circle,fill=black,minimum size=10pt,inner sep=0pt]
\tikzstyle{circled} = [draw, blue, thick, circle, minimum size=.55cm]
\tikzstyle{edge}=[<-,color=black,line width=1.5]
\tikzstyle{edgeback}=[->,color=black,line width=1.5]
\tikzstyle{edgelabel}=[color=black]
\begin{tikzpicture}[scale=0.8, auto]
    \node[vertex] (c0) at (0,0) [label=left:$v$]{};
    \node[vertex] (c1) at (1,0) [label=below left:$x$]{}
        edge[color=black,line width=1.5] node {} (c0);
    \node[vertex] (c2) at (2,0) {};
    \node[vertex] (c3) at (3,0) [label=below right:$s$]{}
   	edge[color=black,line width=1.5] node {} (c2);
    \node[vertex] (c_end) at (6,0) [label=right:$t$]{};
    

    \draw [decorate,decoration={brace,amplitude=8pt},xshift=0pt,yshift=8pt, line width = 1]
    (0,0) -- (3,0) node [black,midway,yshift=8pt] {$\ell/2$};
    \draw [decorate,decoration={brace,amplitude=8pt},xshift=0pt,yshift=-8pt, line width = 1]
    (3,0) -- (1,0) node [black,midway,yshift=-8pt] {$k$};

    \node[draw=none] at (1.525,0) {$\cdots$};
    \node[draw=none] at (3.525,0) {$\cdots$};
    \node[draw, black, thick, rectangle] at (4.525,0) {$H_{\ell/2}$};
    \node[draw=none] at (5.525,0) {$\cdots$};
\end{tikzpicture}
\caption{$H$, which consists of a simple path of length $\ell/2$, whose vertex $s$ is the vertex $s$ of in-degree $0$ in $H_{\ell/2}$, depicted in Figure~\ref{fig:dist_big_prob}.}
\label{fig:dist_is_bad}
\end{figure}

\begin{prop} The algorithm that minimizes the maximum distance from a seed has approximation ratio $O(\sqrt{n}/2^{n/6})$ when $\alpha=1/2$. 
\end{prop}
\begin{proof}
Suppose we can choose at most one seed in $H$, shown in Figure~\ref{fig:dist_is_bad}.  Minimizing the max distance means the seed we use is $s$, and for sufficiently large $\ell$ the minimum probability is $p_v = \alpha^{\ell/2}$, at least for $\alpha = 1/2$ (using the previous lemma).  However, the optimal seed to use is $x$, where $x$ is a vertex $k\le \ell/2$ distance from $s$.  Under this seed set, $p_v$ remains the vertex with the minimum probability of getting infected so long as, for some constant $c$, $\alpha^{\ell/2-k} \le \frac{2c\alpha^{k}}{\ell}$ (again using the previous lemma).  Solving for $k$ to maximize the minimum probability, we get $k = \ell/4 - \frac{\log\left(\frac{\ell}{2c}\right)}{2\log\left(\frac{1}{\alpha}\right)}$.  Then the approximation ratio is no better than
$\frac{\alpha^{\ell/2+1}}{\alpha^{\ell/2-k+1}} = \alpha^k = \frac{\sqrt{\ell}\alpha^{\ell/4}}{\sqrt{2c}} =O(\sqrt{\ell} (1/2)^{\ell/4})$,
and finally note $H$ has $\frac{5}{2}\ell-4$ edges, $\frac{3}{2}\ell+1$ vertices, and the maximum in-degree (and out-degree) is two.
\end{proof}

Despite the hardness results of this section, we will show in the next section that these algorithms perform well in practice.




\section{Experiments}
\label{sec:experiments}

\begin{figure*}
  \centering
\begin{subfigure}[t]{.33\textwidth}
  \includegraphics[width=\columnwidth]{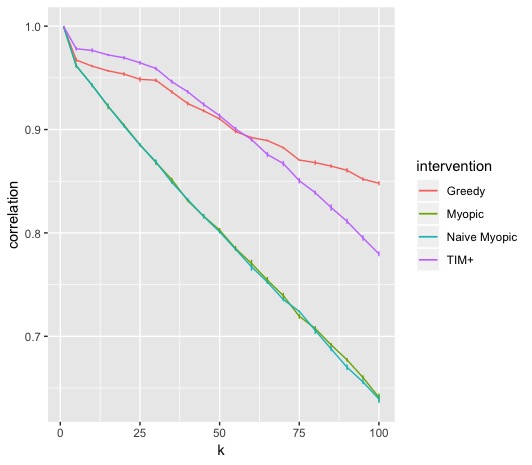}
\caption{Initial probabilities (with one seed).}
\label{fig:arxiv_init_probs_spearman} 
\end{subfigure}
\begin{subfigure}[t]{.33\textwidth}
  \centering
  \includegraphics[width=\columnwidth]{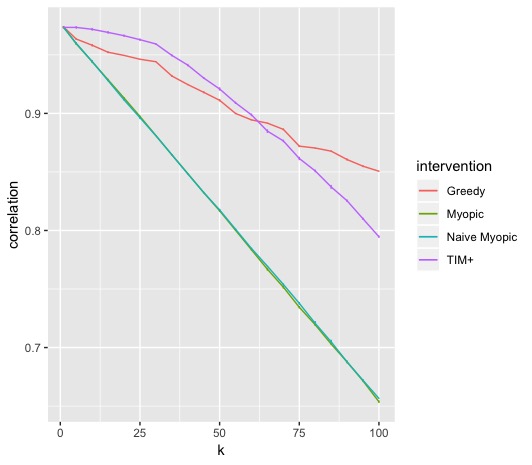}
\caption{Degree of the vertex.}
\label{fig:arxiv_deg_spearman} 
\end{subfigure}
\begin{subfigure}[t]{.33\textwidth}
  \centering
  \includegraphics[width=\columnwidth]{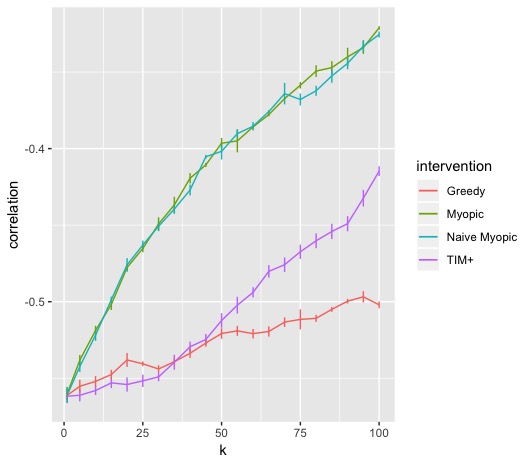}
\caption{Distance from the center.}
\label{fig:arxiv_dists_spearman} 
\end{subfigure}
\caption{Correlations between the set of probabilities of access after intervention and three proxies for position in a network in the Arenas graph.  Bars correspond to one standard deviation computed over 20 runs of each of the heuristics.}
\label{fig:spearman_corrs}
\end{figure*}

Our experimental evaluation will investigate the following question:  does
maximizing $\mu_{-\infty}$ create real changes in access? Is this different from
the interventions achieved via maximum reach? And how effective are the proposed
strategies for optimizing $\mu_{-\infty}$? Since our goal in this paper is to
introduce and validate a method for reducing access gaps, we will not focus on
achieving the fastest implementations (although we will compare the efficiency
of different heuristics).

\subsection{Experimental procedure}
\label{sec:data-sets}

For our evaluation, we used social networks sourced from the SNAP~\cite{snapnets}
and ICON~\cite{icon} repositories as described in Table~\ref{table:data_sets}.  

$\mu_{-\infty}$ is a stringent objective function:  it minimally requires having at least one seed in every connected component to achieve non-zero minimum probability, which may require a large number of added seeds if for example there are many disconnected nodes.  Since the access gap is maximally large if there is at least one seed and a vertex with $p_v=0$, we assume that the number of added seeds is large enough to cover all components of the graph.  This allows us to add seeds to each component of the graph separately.  As a simplifying assumption, in the experiments, we only consider the case (in directed graphs) when the components are strongly connected.  In particular, rather than running the heuristics on all of the components, we just use the largest strongly connected component of the graph.

We also varied our intervention size between $k=1$ and $100$, independent of the
size of the graph. This is a typical number of seeds used for interventions in
the literature, and considering the application -- recommending a job position -- is a
practical intervention size. We varied $\alpha$ -- the probability of message
transmission across an edge -- in the range $\left \{ 0.1, 0.2, 0.3, 0.4, 0.5
\right \}$\footnote{We report results for $\alpha \ge 0.3$ for
  brevity. Behavior below this range was similar.}. Above this range information spreads so effectively that all
algorithms are indistinguishable. Below this range the utilities $p_v$ obtained
are small enough that it is hard for Monte Carlo estimation to distinguish between
them. We run $1000$ simulations in order to estimate probabilities for any given seed set and repeat each heuristic $20$
times, reporting the average result.

As a baseline, we used the algorithm TIM+~\cite{tang_influence_2014}, which was designed to optimize maximum reach. While this procedure is not a true baseline (it does not directly optimize $\min p_i$), it provides insight into
how existing methods for maximum reach might work in this newer setting.
We also use as a baseline picking the $k$ seeds uniformly at random (which we will refer to as \rand). 

\begin{table}
\begin{tabular}{ p{2cm} p{1.5cm} p{1.5cm} p{1.5cm}  }
 \toprule
  Name & Nodes & Edges & Direction\\
  \midrule
\href{http://snap.stanford.edu/data/email-EuAll.html}{EU}~\cite{eu} & 803 & 24729 & Directed\\
\href{http://konect.uni-koblenz.de/networks/arenas-email}{Arenas}~\cite{konect:2017:arenas-email,konect:guimera03} & 1133 & 5451& Directed\\
\href{https://toreopsahl.com/datasets/}{Irvine}~\cite{irvine} & 1294 & 19026 & Directed\\
\href{http://snap.stanford.edu/data/ego-Facebook.html}{Facebook}~\cite{facebook} & 4039 & 24729 & Undirected\\
\href{http://snap.stanford.edu/data/ca-GrQc.html}{ca-GrQc}~\cite{eu} & 4158 & 13428 & Undirected\\
\href{http://snap.stanford.edu/data/ca-HepTh.html}{ca-HepTh}~\cite{eu} & 8638 & 24827 & Undirected\\
 \bottomrule
\end{tabular}
\caption{Overview of the data sets we use. \label{table:data_sets}}
\vspace*{-0.2in}
\end{table}

\subsection{Maximin and network structure}
\label{sec:max-reach-reflects}

In practice, what are the effects of using maximin over max reach as the
objective?  We give evidence that when maximizing reach instead of using
maximin, interventions end up strongly reflecting the existing structure of the
network.  That is, vertices are more likely to become seeds if they are close to
the center of the network, where probabilities of receiving the information are
\emph{already} high and do not need as many additional interventions.

We show this by measuring the correlation between the probability of receiving
information before intervention versus after intervention.  We use as a simple proxy
for `before intervention' the probabilities $p_v$ when the vertex with the highest degree is the sole seed.
Figure~\ref{fig:arxiv_init_probs_spearman} shows the correlation between these
two sets of probabilities in the Arenas graph, and indeed the correlation is
significantly higher when using TIM+ than when using \greed.

Assuming every vertex is equally deserving of information, we do not want
`well-positioned' vertices to have an advantage simply because they are
well-positioned.  Thus, we look at the correlation between the probability of
information access after intervention and a few other proxies for position in a network.
Figures~\ref{fig:arxiv_deg_spearman} and~\ref{fig:arxiv_dists_spearman} show the
results for the degrees of the vertices as well as their distances from the
center of the graph.  Using TIM+, as the distance decreases towards the center
or the degree of the node increases, the probabilities of information access increase,
leading to a larger (negative) correlation.  Again, this effect is lessened by
using \greed, whose resulting probabilities correlate less than TIM+ with both
the degree of the vertex and the distance from the center.  In other words, \greed~reduces the correlation between vertices' probability of receiving the information and how well connected the vertices are. \nvgreed~yields very similar results to \greed, as again seen in Figure~\ref{fig:spearman_corrs}.

\begin{figure}[htbp]
  \centering
  \includegraphics[width=0.81\columnwidth]{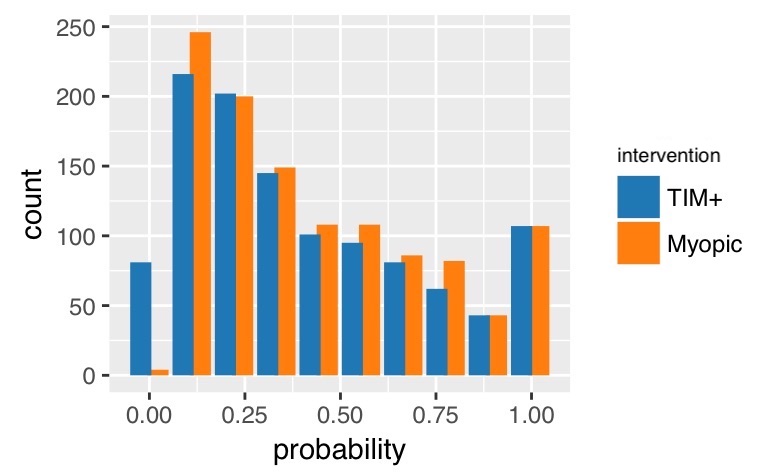}
  \caption{Distribution of probabilities over all vertices in the Arenas graph after adding 100 seeds with $\alpha = 0.1$.}
  \label{fig:histogram}
\end{figure}

In addition, \greed\ changes the distribution of probabilities $\{p_{v_1},\ldots,p_{v_{|V|}}\}$.  Not only does it decrease the number of vertices with very low probability of receiving the information, but it also increases the number of vertices with larger probabilities over a broad range of probabilities, as seen in Figure~\ref{fig:histogram}.

\begin{figure*}
  \centering
\begin{subfigure}{.49\textwidth}
  \includegraphics[width=\columnwidth]{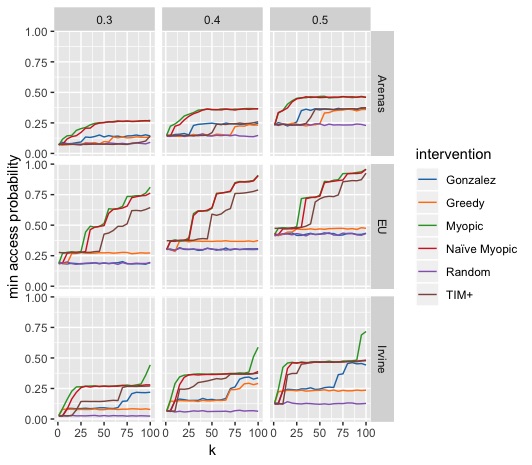}
\caption{Smaller data sets.}\label{fig:smaller}
\end{subfigure}
\begin{subfigure}{.49\textwidth}
  \centering
  \includegraphics[width=\columnwidth]{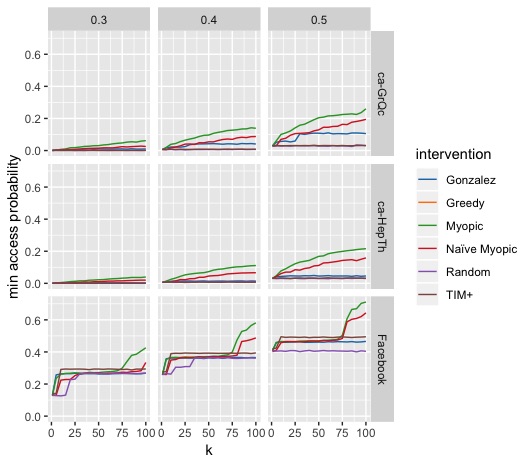}
\caption{Larger data sets.}\label{fig:larger}
\end{subfigure}
\caption{Comparison of the six heuristics with respect to the minimum probability for values of $\alpha = \{0.3, 0.4, 0.5\}$.}\label{fig:heuristic_comp}
\end{figure*}

%
%

\subsection{Heuristic performance}
\label{sec:heuristics-max-min}

We now study the behavior of the heuristics described in the previous
section. We would like to know how they compare in terms of effectiveness
(maximizing $\mu_{-\infty}$) and speed.

We present effectiveness results in
Figure~\ref{fig:heuristic_comp}. We omitted the heuristic \ahead~when
experimenting with larger data sets because it was prohibitively slow. Note that
in both charts, the \greed~and \nvgreed~heuristics consistently outperform the
other methods for all ranges of $\alpha$ and intervention size $k$. The
heuristics that do not use estimation are all consistently poor performers, and TIM+ performs well but is
consistently dominated. For the smaller data sets, shown in Figure~\ref{fig:smaller}, \ahead~also does fairly well.

%

The running time of the heuristics is summarized in
Table~\ref{table:time}, which shows there is a natural tradeoff
between running time and effectiveness. In
particular, while the methods that make no use of estimation yield poorer quality
results, they run extremely fast because they avoid the expensive step of
estimating probabilities. Among the heuristics that estimate probabilities, \nvgreed~is
the fastest, with TIM+ also comparable, while the
\greed~heuristic is an order of magnitude more expensive.  \ahead~is
still another order of magnitude slower than \greed, making it prohibitively expensive to compute in even relatively small graphs.

\begin{table}
\begin{tabular}{cccc}
\toprule
Algorithm & \multicolumn{3}{c}{Average time (s)} \\
& Arenas & EU & Irvine \\
\midrule
\rand & 0.007 & 0.015 & 0.012 \\
\gonz & 0.021 & 0.031 & 0.033 \\
\nvgreed & 0.086 & 0.208 & 0.184 \\
TIM+ & 0.876 & 1.826 & 1.046 \\
\greed & 8.910 & 19.438 & 16.755 \\
\ahead & 507.35 & 759.296 & 1399.26\\
\bottomrule
\end{tabular}
\caption{Speed of each of the heuristics on three data sets for 100 seeds.  Times to completion are averaged over 20 runs\label{table:time}.}
\vspace*{-0.2in}
\end{table}


\subsection{Performance on max reach}
\label{sec:performance-max-reach}

\begin{figure}
\centering
\includegraphics[width=\columnwidth]{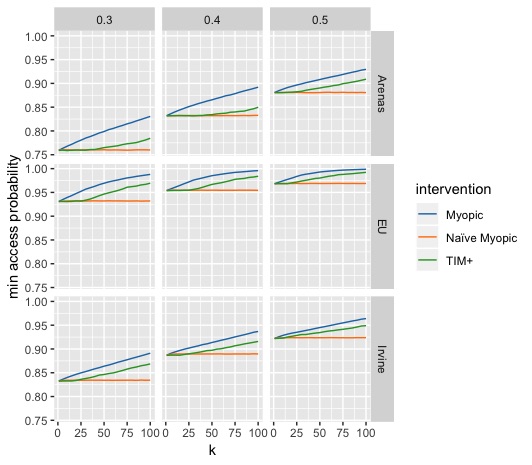}
\caption{Comparison of three heuristics with respect to reach, the average probability after intervention, for $\alpha = \{0.3, 0.4, 0.5\}$.}
\label{fig:reach}
\vspace*{-0.05in}
\end{figure}

While the goal of the introduced heuristics is to maximize the minimum
information access, it is also valuable to measure them by their average reach~$\frac{1}{|V|}\sum_{v\in V} p_v$ to see if they are effective at spreading information to a large number of vertices.  We compare the performance of \nvgreed~and \greed~to TIM+ on this measure over three datasets 
(see Figure~\ref{fig:reach}).  The results show that while \nvgreed\ does not
perform well to maximize reach, \greed\  appears to outperform TIM+ even
though TIM+ was designed for average reach and \greed\ was
not.  This is likely because each seed added by \greed~ is guaranteed to increase
reach on the graphs, while algorithms that focus on maximizing reach may
inadvertently provide access to nodes already reached.  However, recall that
\greed\ is much slower than TIM+ (see Table~\ref{table:time}) and so this
potential improvement does not come without pitfalls. This tradeoff
between average and minimum reach seems worthy of further study.

\clearpage

\bibliographystyle{ACM-Reference-Format}
\balance
\bibliography{main}

\end{document}